\documentclass[aps,prb, twocolumn, showpacs]{Tsurumaki}
\usepackage{graphicx}
\begin{document}


\title{Superconducting weak link generated on a linear array of substrate steps in ultrathin films of YBa$_2$Cu$_3$O$_{7-\delta}$}


\author{A. Tsurumaki}
\author{Z. Hiroi}
\affiliation{Institute for Solid State Physics (ISSP), The University of Tokyo, Kashiwa, Chiba 277-8581, Japan}


\date{\today}

\begin{abstract}
We have studied a three-unit-cell-thick YBa$_2$Cu$_3$O$_{7-\delta}$ (YBCO) superconducting film grown on a slightly vicinal SrTiO$_3$ substrate that was polished from the (001) plane by a small miscut angle (less than 0.5$^\circ$). This produced, on its surface, a linear array of single-unit-cell-height steps with a period larger than 40 nm. A characteristic fault, with a displacement corresponding to one-third of the {\itshape c}-axis length of YBCO, is generated in the film on each substrate step, as observed by atomic force microscopy. Transport measurements with current flows parallel and perpendicular to the fault show that the fault serves as a superconducting weak link, as well as a unique one-dimensional flux-pinning center in an applied magnetic field. In addition, a characteristic fine structure is found in the flux-flow resistivity as a function of magnetic field, the origin of which may be ascribed to a matching effect between the periodically aligned faults and the flux lattice.
\end{abstract}


\maketitle

Extensive studies of various transition metal oxides have been carried out over the last decade using high-quality epitaxial thin films. Many interesting phenomena have been found, which only manifest themselves in thin films \cite{1}. In order to obtain a high-quality thin film, the surface of the substrate materials is absolutely important. For the (001) surface of SrTiO$_3$ (STO), which is the most typical substrate material, a method to obtain a nearly ideal surface with atomically flat terraces and steps has been established utilizing NH$_4$F-HF (BHF) etching \cite{2} and high-temperature annealing \cite{3}. However, there still remain single-unit-cell-height steps on the surface, which must be taken into account when one studies the property of an ultrathin film.

\begin{figure} [t]
\centering
\includegraphics[width=7.9cm, clip]{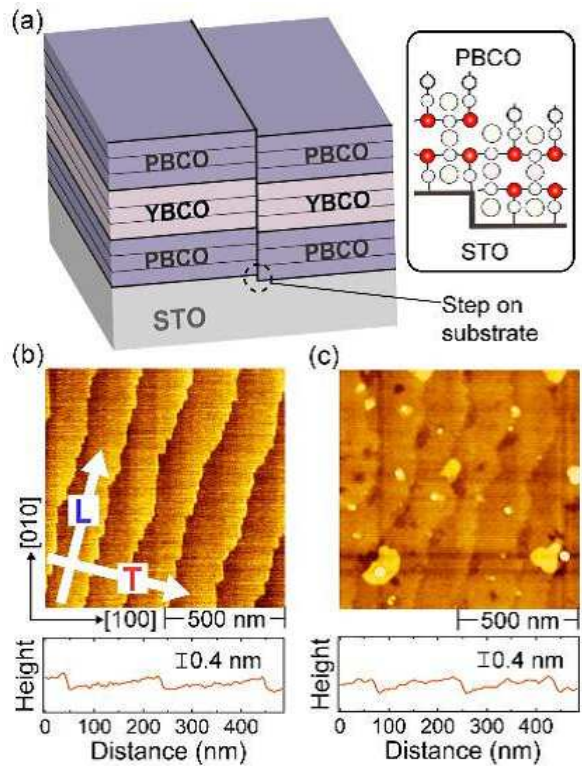}
\caption{\label{fig1}(a) Schematic drawing of a sample consisting of a three-unit-cell-thick YBCO layer sandwiched between two three-unit-cell-thick PBCO layers grown on a STO substrate with a periodic array of one-unit-cell-height steps. AFM images of the surfaces of a 0.1$^\circ$ miscut STO substrate and the first three-unit-cell PBCO layer are shown in (b) and (c), respectively. Directions parallel and perpendicular to the step line are denoted by {\itshape L} and {\itshape T}, respectively. A surface profile along the {\itshape T} direction is shown below each image.}
\end{figure}

In this work we focus on a cupric oxide superconductor YBa$_2$Cu$_3$O$_{7-\delta}$ (YBCO) having essentially a two-dimensional electronic state. Thus far, a number of studies on YBCO in the form of ultrathin films have been performed \cite{4, 5, 6, 7}. For example, Matsuda {\itshape et al}. reported that the zero-resistivity temperature {\itshape T}$_{c0}$ of ultrathin YBCO films at the superconducting transition decreased with a decrease of the numbers of unit cell {\itshape n}; thus, {\itshape T}$_{c0}$ = 80, 63 and 30 K for {\itshape n} = 10, 3 and 1, respectively \cite{6}. This reduction of {\itshape T}$_{c0}$ has been explained by assuming the presence of a Kosterlitz-Thouless (KT) transition \cite{5, 6}, or otherwise a depletion of carrier density by interfacial charge transfer \cite{7}, neither of which consider possible influences from the substrate step. Since the {\itshape c}-axis length of YBCO (1.17 nm) is approximately triple that of STO (0.39 nm), the substrate step must generate a ``fault" in a {\itshape c}-axis oriented YBCO film. As schematically depicted in Fig. \ref{fig1}(a), only one of the pair of CuO$_2$ planes in the unit cell can be connected with another plane terminating at the fault, and, thus, its influence on the superconducting properties must be serious for an ultrathin YBCO film.

On the other hand, several recent studies on the flux pinning in YBCO films have focused on the influence of the substrate step \cite{8, 9, 10, 11, 12}. They used vicinal STO substrates having large miscut angles of $\theta$ = 1 $\sim$ 8$^\circ$ off the (001) plane or the (106) surface ($\theta$ = 9.5$^\circ$). On the surface of these substrates, there are linear arrays of closely-spaced steps at intervals of 3 $\sim$ 18 nm. In thick YBCO films grown on them, structural faults generated by the substrate steps have been actually observed by means of cross-sectional transmission electron microscopy; these are called antiphase boundaries (APBs) \cite{8, 9, 10}. Interestingly, anisotropy in the critical current density, which depends on whether it is measured parallel or perpendicular to the step line, has been observed and ascribed to the anisotropic flux pinning caused by the APBs, as evidenced mostly by magneto-optical measurements \cite{8, 9, 10, 11, 12}. However, because the high-density steps often cause deterioration of the grown films \cite{8, 12}, the true effect of the substrate steps may be obscured for such vicinal substrates. Moreover, macrosteps that have heights corresponding to multiple unit cells of STO \cite{10} exist on the surface of highly vicinal substrates. Thus, in order to investigate the influence of single-unit-cell-height substrate steps, it is necessary to use a substrate with a smaller miscut angle, which gives less step lines separated by a wider terrace on the surface. On the one hand, it has been reported that faults generated by the substrate steps tend to disappear as the film thickness increases above a few nanometers \cite{8, 9}. Therefore, the preparation of an ultrathin YBCO film is another key issue.

In this report, we have studied high-quality, ultrathin YBCO films grown on slightly vicinal STO (001) substrates, with miscut angles less than 0.5$^\circ$, to search for step-induced phenomena. A characteristic one-third structural fault is generated on each substrate step in the YBCO film, which gives rise to a superconducting weak link. In addition, a matching effect between the periodically aligned one-dimensional faults and the flux lattice is observed for flux-flow resistivity in applied magnetic fields.

Samples were prepared on miscut STO (001) substrates by the pulsed laser deposition method at a substrate temperature of 670$^\circ$C and in an oxygen pressure of 1.5 $\times$ 10$^{-2}$ torr including 9$\%$ of ozone. They consisted of three layers, Pr$_{1.1}$Ba$_{1.9}$Cu$_3$O$_{7-\delta}$ (PBCO) / YBCO / PBCO, with each having a thickness corresponding to three unit cells, as depicted schematically in Fig. \ref{fig1}(a). PBCO is a non-superconductor having the same crystal structure as YBCO and was used to provide buffer and cap layers, to reduce the strain in the YBCO layer from the substrate and to protect the surface of the YBCO layer. We used several STO substrates polished off the (001) plane approximately toward the [100] direction with miscut angles $\theta$ of 0.1 $\sim$ 0.5$^\circ$ (Shinkosha Co., Ltd.). Before the film growth, they were etched in a BHF solution and annealed at 820 $\sim$ 900$^\circ$C in 10$^{-6}$ torr O$_2$. The thickness of films grown was controlled by observing an oscillation in the reflection intensity from high-energy electron diffraction. The epitaxial {\itshape c}-axis oriented growth was confirmed by x-ray diffraction measurements.

Atomic force microscopy (AFM) has observed a quasi-periodic linear array of single-unit-cell-height steps on the surface of a clean STO substrate, typically as shown in Fig. \ref{fig1}(b). The average spacing of steps {\itshape l} is approximately 200 nm, which is close to the value 220 nm expected from $\theta$ = 0.1$^\circ$. In the case of a substrate with $\theta$ = 0.5$^\circ$, {\itshape l} was found to be 48 nm close to the expected value of 45 nm. All the steps have a height corresponding to a single unit cell, 0.39 nm, with no macrosteps observed. Even after the deposition of a three-unit-cell thick PBCO film on the STO substrate, a similar array of steps with nearly equal periodicity is observed, as shown in the AFM image of Fig. \ref{fig1}(c). Since the step height is the same as the pre-existing one on the STO substrate and is obviously smaller than the unit cell of PBCO, a one-third fault must have been generated within the PBCO layer on each substrate step, as schematically depicted in Fig. \ref{fig1}(a). Such 0.4 nm-height steps are also observed on the top surface of a tri-layered sample containing a total of nine unit cells of YBCO and PBCO, though the appearance in the AFM image becomes less clear. When the total film thickness is larger than $\sim$10 unit cells, other types of steps appear with a height corresponding to the unit cell of YBCO or PBCO. This means that, as the film thickness increases, the memory of the substrate step tends to be lost by the introduction of the extra atomic planes. Thus, the one-third fault can exist only in an ultrathin YBCO film.

Electrical resistivity measurements were carried out by the standard four-probe method in two directions, a longitudinal ({\itshape L}) direction along the step line and a transverse ({\itshape T}) direction perpendicular to the step line [Fig. \ref{fig1}(b)]. A sample was cut into a bar, approximately 0.7 mm $\times$ 3.3 mm, with the longer axis along either the {\itshape L} or {\itshape T} direction. Resistivity values were calculated by assuming a current flow only within the middle YBCO layer. This may be reasonable at low temperatures (below 100 K), because the resistivity of metallic YBCO is much lower than semiconducting PBCO. At high temperatures, however, this underestimates the resistivity of YBCO.

\begin{figure} [t]
\centering
\includegraphics[width=7.8cm, clip]{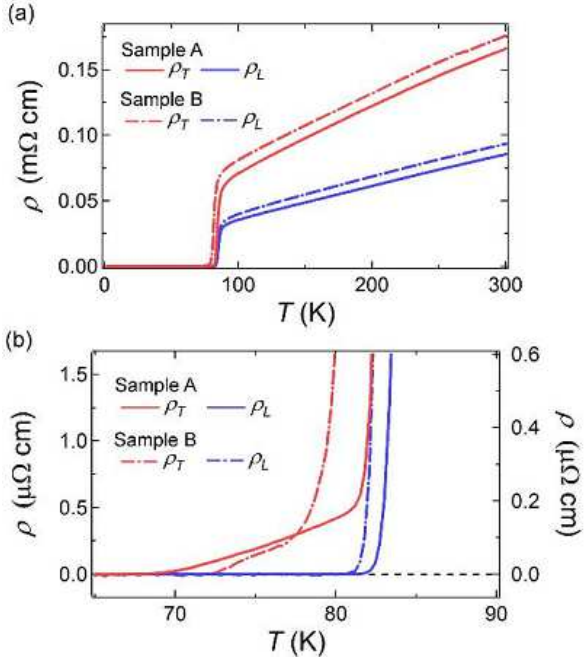}
\caption{\label{fig2}Temperature dependence of the resistivity of two samples, A and B, prepared using the same conditions, on 0.5$^\circ$ miscut substrates measured along the {\itshape L} and {\itshape T} directions. The lower temperature part of transition around {\itshape T}$_{c0}$ is enlarged in (b). An excitation current of 5 $\mu$A was used.}
\end{figure}

Figure \ref{fig2} shows the temperature dependence of the resistivity of two samples A and B grown under the same conditions on STO substrates with $\theta$ = 0.5$^\circ$ ({\itshape l} = 48 nm) in the {\itshape L} and {\itshape T} directions. The two samples exhibit similar behavior, illustrating good reproducibility of the data. The onset of the superconducting transition is approximately 93 K for each sample and in each direction; the midpoint is 85.2 K (83.4 K) in the {\itshape L} direction and 84.8 K (81.9 K) in the {\itshape T} direction for sample A (B). Thus, there is almost no difference in {\itshape T}$_c$ between the two directions in each sample. The transition widths defined by 10 - 90$\%$ of the resistivity drop are also similar, about 6.0 K.

In sharp contrast, there is a marked difference near the {\itshape T}$_{c0}$ between the two directions, as shown in Fig. \ref{fig2}(b). In sample A, the resistivity reaches zero at 81 K for {\itshape L}, while, for {\itshape T}, a long tail appears down to {\itshape T}$_{c0}$ = 66 K, 15 K lower than for {\itshape L}. This is obviously due to the effects of the substrate step. Faults generated on the step probably act as superconducting weak links only for a current running perpendicular to the step line. In sample B, zero resistivity is attained at 80.5 K for {\itshape L} and 71.6 K for {\itshape T}. Thus, a resistivity tail is also present, but the shape of the tail is slightly different from that of sample A. We have observed similar tails for five samples grown on substrates with $\theta$ = 0.5$^\circ$, but only in the {\itshape T} direction.

In a previous study performed without taking care of the existence of substrate steps \cite{5,6} the {\itshape T}$_{c0}$ of a three-unit-cell-thick YBCO film was reported to be 63 K, and the transition width was about 15 K. The {\itshape T}$_c$ onset ($\sim$80 K) and {\itshape T}$_{c0}$ (63 K) in their sample coincide with our {\itshape T}$_{c0}$ values for the {\itshape L} and {\itshape T} directions, respectively. This suggests that the low {\itshape T}$_{c0}$ value is due to the influence of the substrate step. Since there will have existed randomly oriented steps on the polished surface of STO in the absence of an intentional miscut, the current always passed through step-induced faults in any in-plane direction. Therefore, the previously reported large decrease in {\itshape T}$_{c0}$ with decreasing film thickness may not be intrinsic, and its interpretation based on the KT transition should be reconsidered. As we have shown above, the intrinsic {\itshape T}$_{c0}$ value of the three-unit-cell-thick YBCO film is as high as 80 K, as obtained for the {\itshape L} direction free from disturbance by the substrate step.

It is to be noted in Fig. \ref{fig2}(a) that there is anisotropy in the normal-state resistivity between the {\itshape L} and {\itshape T} directions. The anisotropy $\rho_T$ / $\rho_L$ is approximately 2 for both samples A and B at 100 K. We found nearly equal ratios for 12 samples grown on various substrates with {\itshape l} = 42 $\sim$ 277 nm ($\theta$ = 0.1 $\sim$ 0.5$^\circ$), implying that the ratio is independent of {\itshape l}. Thus, this anisotropy must come from the bulk properties. Similar anisotropy of 1.5-5.5 was reported in previous studies for thicker films on substrates with large $\theta$, which was partly attributed to the contribution from the {\itshape c}-axis conduction to the resistivity only in the {\itshape T} direction \cite{8,12}. However, this contribution is negligibly small in the present samples: given the ratio 100 between the resistivities along the {\itshape c}-axis and the {\itshape ab} plane in bulk YBCO, $\rho_T$ / $\rho_L$ is calculated to be 1.008 for $\theta$ = 0.5$^\circ$, using the equation given by Haage {\itshape et al}. \cite{8}. Alternatively, the origin of this resistivity anisotropy may be explained by assuming the preferred orientation of CuO chains along the substrate step. It is known that the {\itshape b}-axis resistivity in YBCO is about half that of the {\itshape a}-axis resistivity because of the contribution of the CuO chains \cite{13}. Moreover, such a preferred orientation has been actually reported in YBCO thick films grown on vicinal substrates with large miscut angles \cite{14,15}.

\begin{figure} [t]
\includegraphics[width=7.5cm, clip]{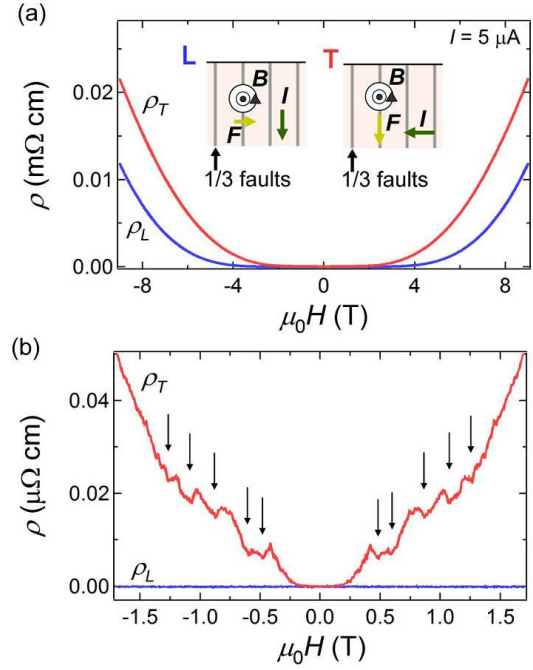}
\caption{\label{fig3}Magnetic-field dependence of resistivity for sample B measured at 70 K. The low-resistivity part is enlarged in (b). Experimental geometries for the {\itshape L} and {\itshape T} measurements are depicted in the inset of (a), where {\itshape I}, {\itshape B} and {\itshape F} denote current, magnetic field and the resulting Lorentz force. Two datasets measured with increasing and decreasing field are shown; they overlap completely. Several dips observed only for {\itshape T} are indicated by arrows in (b).}
\end{figure}

We have noticed further interesting differences in the resistivity between the two directions measured in applied magnetic fields. Figure \ref{fig3} shows the magnetic field dependence of the resistivity in the {\itshape L} and {\itshape T} directions for sample B measured at 70 K, just below {\itshape T}$_{c0}$ = 72 K for the {\itshape T} direction. Magnetic fields are applied perpendicular to the film surface. Zero resistivity is retained for the {\itshape L} direction up to a large magnetic field of 2 T within the experimental resolution, while, for the {\itshape T} direction, finite resistivity is induced by a small field of about 0.1 T. This finite resistivity must be due to a flux-flow resistivity, because the applied fields are small and the measured temperature is close to {\itshape T}$_{c0}$ for the {\itshape T} direction. The observed difference implies that the step-induced fault serves as a one-dimensional flux pinning center; the Lorentz force is perpendicular to the step line for {\itshape L}, but parallel for {\itshape T}, as depicted in the insets of Fig. \ref{fig3}(a). This implication from the transport measurements for an ultrathin YBCO film is consistent with previous results obtained by magneto-optical measurements on thick films grown on substrates with larger miscut angles \cite{8,10,11}.

Remarkably, an additional feature is observed in the flux-flow resistivity measured in the {\itshape T} direction. There are many tiny dips at certain magnetic fields. The noise level in our measurements (5 $\times$ 10$^{-10}$ $\Omega$ cm) is much smaller than these structures. In addition, two datasets measured with increasing and decreasing fields overlap completely with a perfectly symmetric profile for the sign reversal of the magnetic field. The dips remain at the same fields at temperatures between 67 and 73 K, while they become less apparent at increasing temperature and finally disappear above 78 K. Increasing the excitation current also causes no change in the dip positions.

Since the dips are observed only in the flux-flow resistivity regime, the origin must be attributed to an enhancement of flux pinning at certain magnetic fields. The most probable explanation is a matching effect between the flux lattice and pinning centers. In previous experiments, the matching effect has been clearly observed both for conventional superconductors \cite{16} and high temperature superconductors \cite{17} by introducing artificial pinning centers arranged in forms such as a triangular lattice. In the present case, the nature of the pinning centers must be quite different from those studied previously since this is an array of one-dimensional structural faults generated by the substrate step. Thus, such a simple matching effect as in triangular defects should not occur.

A matching effect for a two-dimensional superconductor with one-dimensional pinning centers has been considered theoretically \cite{18}. Given the spacing of the step lines for sample B (48 nm) in the equation given by Martinoli \cite{18}, we estimate possible matching fields for the triangular flux lattice as 111, 194, 259, 345, 444, 777 and 1036 mT. The observed magnetic fields for major dips are 100, 200, 240, 320, 380, 480, 580, 760, 870, 1075 and 1265 mT. Thus, many of them are successfully reproduced. Small differences between the observed and calculated fields may be understood by assuming a spontaneous small deformation of the flux lattice \cite{18}. However, the dips observed at 580, 870 and 1265 mT cannot be explained by this matching effect.

The occurrence of such unaccountable resistivity dips has been also reported for artificial triangular pinning centers. The origin is not clear, but may be ascribed to an enhanced matching related to interstitial flux \cite{16} or fractional matching effects \cite{19}. More complicated matching may be expected in the present case for an array of one-dimensional pinning centers. Alternatively, we have to take into account the fact that our natural system is not as structurally perfect as the artificial systems. The periodicity of steps may have a distribution and meandering of step lines can occur, as actually observed in Fig. \ref{fig1}(b). These inevitable imperfections would make the pinning properties complicated. Therefore, it is necessary to measure transport properties on smaller areas created using a micro fabrication technique, which will be the subject of future work.

In summary, we have introduced a one-dimensional array of one-third faults in three-unit-cell-thick YBCO films by utilizing one-unit-cell-height steps on slightly vicinal STO substrates. It is shown that this fault serves as a superconducting weak link for a current perpendicular to the fault, giving rise to a tail in the resistive transition. Furthermore, we find many dips in the flux-flow resistivity at specific magnetic fields, suggesting a unique matching effect between the linear array of faults and the flux lattice.

We thank Y. Murakoa for his contributions at the initial stage of this work. We also thank K. Hirata for helpful discussion. This work was supported by the Global COE Program ``Physical Sciences Frontier", given by MEXT.


\end{document}